\documentclass[11pt]{article}
\usepackage{epsfig}
\usepackage[usenames]{color}
\usepackage{graphicx}
\usepackage{amsmath}
\def\be{\begin{eqnarray}}\def\ee{\end{eqnarray}}
\def\lsim{\mathrel{\rlap{\lower3pt\hbox{\hskip1pt$\sim$}}
     \raise1pt\hbox{$<$}}} %less than or approx. symbol
\def\gsim{\mathrel{\rlap{\lower3pt\hbox{\hskip1pt$\sim$}}
     \raise1pt\hbox{$>$}}} %greater than or approx. symbol
\def\le{ \begin{array}{ll}}\def\re{\end{array}}

\def\lear{ \left( \begin{array}{cc}}\def\rear{\end{array} \right)}

\def\le{ \left( \begin{array}{cc}}\def\re{\end{array} \right)}

\def\bi{\bibitem}

\def\eft-hls{{\it EFT}$_{\rm bsHLS}$}
\def\skyrmion-hls{{\it Skyrmion}$_{\rm sHLS}$}
\def\Tr{\rm Tr}
\def\del{\partial}
\def\mathbb{\cal A}

\renewcommand{\thefootnote}{\fnsymbol{footnote}}
\begin{document}
\centerline{\Large \bf 
The Smile of Cheshire Cat  At High Density}
\vskip 0.5cm
\begin{center}
	
{Mannque Rho}
\footnote{\sf e-mail:mannque.rho@ipht.fr}

\em{Universit\'e Paris-Saclay, CNRS, CEA, Institut de Physique Th\'eorique, \\ 91191, Gif-sur-Yvette, France }

\vskip 0.3cm
{\ (\today)}
\end{center}
%\maketitle

%\tableofcontents

\centerline{ \bf ABSTRACT}

\noindent  Baryons in finite nuclei, nuclear matter and dense compact-star matter are described in terms of Cheshire Cat for  QCD. A potential conceptual link, admittedly short in mathematical rigor, between their manifestations is  made by what's called Cheshire Cat Principle. Put in terms   ``dual" to QCD variables, going to very high density exposes quantum Hall droplets -- or pancakes --  at which the dilaton-limit fixed-point (DLFP) with $g_A\to 1$, $f_\pi\to f_\chi$ -- where $f_\pi$ and $f_\chi$ are respectively the pion and dilaton decay constants -- and the baryon parity-doubling are reached. This scenario suggests a thus-far totally unexplored structure of dual baryonic matter at high density which does neither require nor  rule out  (rapid) first-order phase transitions from hadrons to quarks in the core of compact stars on the verge of gravitational collapse.


\setcounter{footnote}{0}
\renewcommand{\thefootnote}{\arabic{footnote}}
\vskip 1.0cm

%\maketitle

\tableofcontents

\section{Introduction}
With the advent of gravity-wave observations with neutron-rich compact stars that are being -- and will be -- discovered, a thus-far more or less unexplored domain of physics is opening up,  resurrecting the old question raised more than a half a century ago: What is the structure of dense baryonic matter stable against, but on the verge of gravitational collapse into black a black hole? Of course general relativity enters with its own multifacets, but what has been baffling in modern physics  is how are the baryons crunched into such high density in the standard model (SM) of particle physics? This question arises in specific terms of what is the equation of state (EoS) of the matter that figures in the TOV (Tolman-Oppenheimer-Volcoff)  equation to describe the properties of massive compact (predominantly neutron) stars stable against gravitational collapse? This question becomes highly topical as the on-going -- and up-coming -- gravitational wave measurements are to provide ultimately precision data on the properties of such massive stars.

\begin{figure}[h]\centering 
\includegraphics[scale=0.2,angle=0]{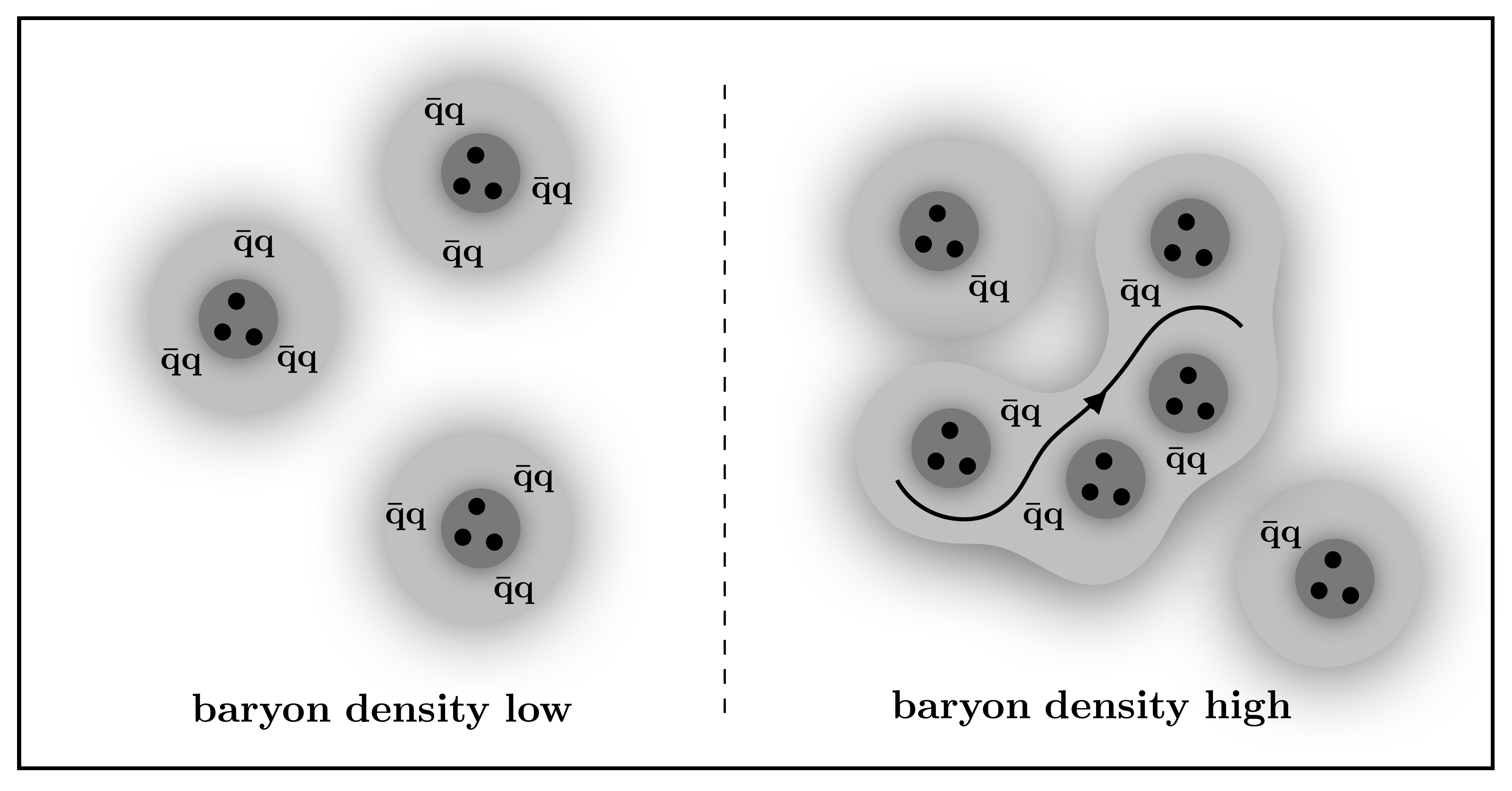}
\caption{Nucleons in nuclear matter described by chiral bags containing three colored quarks, surrounded by meson cloud ($\pi$ plus heavier mesons). Illustration: Courtesy from \cite{weise}. Left panel: low-density matter, right panel: high-density matter. Chiral bags increase in size as density increases.}
\label{Weise1}
\end{figure}

This question brings back the long-standing issue on the basic structure of nucleon in the fundamental theory of strong interactions  (now belonging to  the Standard Model),  widely discussed several decades ago, as a model of QCD,  the  MIT bag model.  There was a big dispute in early 1970's between  Stony Brook, Long Island and the nuclear and particle theorists at MIT in Cambridge where the ``MIT Bag"  was conceived~\cite{MITbag} on the structure of baryons in heavy nuclei.  Taking  the well-studied heavy nucleus $^{208}$Pb as a specific example, the MIT theory argued that  this nucleus should be more properly described in terms of $3\times 208=624$  quarks instead of 208 nucleons given that the MIT bag for the single nucleon was to have the confinement size  $R_{conf} \sim 1$ fm. This would depict the nucleon bags more or less close packed in heavy nuclei, giving the picture of the nucleons resembling  ``grapefruits in a salad bowl." A nucleon has the size of radius $\sim 1$ fm and a nucleon in the MIT bag model contains $N_c=3$ quarks {\it confined} in the bag with a confinement radius $\sim 1$ fm. This structure was considered,  generally at the time,  in nuclear community and particularly by the Stony Brook theorists,   ``glaringly" at odds with the then generally accepted description of  nucleons in the Pb nucleus treated in the shell model with the nucleons pictured as quasi-fermions moving in  an average potential.  The nucleon was considered as a small object of radius much less than 1 fm  surrounded by meson cloud with energy much larger than the nucleon mass $m_N \sim 1 {\rm GeV}$,  in consistency with the old Yukawa theory of nuclear forces with strong self-energy processes with the pion field -- and subsequently heavier meson fields --  bringing down the nucleon mass to its known measured value by self-energy corrections.  Following the notion of bag-confinement  in QCD,  while I was spending a year at the Nuclear Theory Group with  Gerry Brown, he and I proposed ``The Little Bag" model~\cite{LB} to contrary the Big MIT bag.  The little bag was depicted as confining three quarks,  massless due to large density restoring chiral symmetry within the bag.  During the whole  period of my stay I was witnessing  the daily on-going disputes taking place at Stony Brook with its Little Bag (LB) and MIT with its Big Bag (BB)\footnote{Gerry used to commute, quite frequently, between Stony Brook and Cambridge giving talks, generating heated arguments with the local protagonists of the BB. To our amusement each time he returned from MIT, his little bag got a little bit bigger, visibly persuaded by the MIT theorists. But it shrunk back after a little while,  dissuaded by us in the Stony Brook group. At his office door Gerry had hung two brown bags, one a big (brown shopping)  bag and another a small (lunch) bag contained inside the big one, clearly indicating his ambivalence.}. This -- what one might call -- ``dichotomy" between the Stony Brook LB and the MIT BB became more striking at higher densities probed in compact stars as illustrated in Fig.~\ref{Weise1} copied from  \cite{weise} where it was used to argue, rather persuasively, against rapid first-order phase transitions from baryons to quarks inside massive compact stars.

The LB was constructed for the nucleons with three colored quarks coupled to the small-mass pions, recognized as the Nambu-Goldstone (NG) bosons resulting from spontaneous breaking of the nearly perfect chiral symmetry. In proceeding with the LB, the quark-pion coupling at the bag boundary was constrained to be consistent  with chiral symmetry with the quarks confined inside the bag and the pions clouded outside.   
With a suitable coupling boundary condition imposed at a given ``confinement" radius $R_{bag}$,  the ``Chiral Bag Model" (CBM)~\cite{CBM} was applied with some success to certain class of phenomena of nuclear dynamics~\cite{CBMapplied}.   But there was however an obvious technical difficulty  due to its too big a size of  the self-energy corrections required to bring down the mass to near its physical value.  

%%%
The way out of this stumbling block came with what one could  say a striking ``tongue-in-cheek" idea proposed by Nadkarni, Nielsen and Zahed  in 1985~ \cite{CCP,TCCM}, inspired from the Cheshire Cat in  Lewis Carroll's  1865 fable  ``Alice in the Wonderland."  The {\it baryon charge} that leaks out from the bag is linked to the {\it smile} that disappears from the face of the Cheshire Cat in the Wonderland.
%%%%

What's  magical in the Cheshire Cat (CC) solution  was  that the nucleon could be described as a soliton, nowadays referred to as skyrmion, discovered by Skyrme in early 1960s~\cite{skyrme}. What happens is that the baryon charge  $B=1$ for a single nucleon can leak out of the LB into the pion cloud residing outside of the bag. The crucial point, elaborated below, is that this leakage is caused by a quantum anomaly lodged at the bag boundary that makes the baryon charge disappear from the bag. This anomaly, associated with a gauge symmetry,  turns out to make the  bag radius $R_{bag}$, or equivalently confinement radius $R_{conf}$, an unphysical quantity. This is the emergence of what is referred to as ``Cheshire Cat Principle (CCP)" and the associated mechanism  ``Cheshire Cat Mechanism (CCM)."  As will be mentioned below there could be what one might view as an alternative approach with the quark-pion coupling made in the whole space invoking {\it no boundary conditions}. This approach, referred to as ``Cloudy Bag Model"~\cite{CloudyBM}, could be considered as a variant of CCM without bag boundary conditions and hence involves no leakage of baryon charge. In this model the baryon charge is strictly confined inside the bag. If the bag radius is taken sufficiently large,  it could be formulated as a particular {\it gauge-fixed version} as described in Sec.~\ref{CCgauge}.

I first recount, qualitatively,  what took place when the CCM was introduced into the LB and how the resulting CC model fared with Nature in nuclear processes. I then describe how  a conceptually new development took place in the efforts to access the properties of superdense neutron matter relevant to massive compact stars currently being observed in gravity-wave and associated measurements. 

\section{The Cheshire Cat}
Reconciling  the ``grapefruits-in-the-salad" dilemma in heavy nuclei, say, $^{208}$Pb in the LB scenario turns out to come from how  topology figures in the baryon structure: It turns out that the BBs in $^{208}$Pb can be shrunk to the LBs relying on that  baryons are  topological solitons. The nucleons, proton and neutron, belonging to the baryon flavor  symmetry $SU(N_f )$ with $N_f=2$, can be considered as skyrmions in the large $N_c$ limit  of QCD~\cite{Witten-qcd}  where $N_c$ is the number of colors, thereby resurrecting Skyrme's idea left dormant for $\sim 2$ decades. The outcome was that the whole baryon octet, not just proton and neutron, can be related to the meson octet via topology.  This is stated mathematically in terms of  the homotopy group $\pi_{3} (SU(N_f))=\cal{Z}$.  But it holds only if $N_f >1$,  not for $N_f=1$. There is no-go theorem for  a single flavor, i.e., $\pi_3 (U(1))=0$, so there can be no flavor-singlet  skyrmion coming from the flavor singlet meson, $\eta^\prime$.  Surprisingly this seems to be due to the color symmetry of  QCD~\cite{noB0}. It turns out the flavor singlet soliton can actually be present topologically but not as a skyrmion. We will come back to this issue which is one of what could constitute a surprising new development in nuclear physics. It leads to a fractional quantum Hall droplet involving Chern-Simons topology. It  seems to play no significant role at low density, but, I suggest,  can figure importantly at high density.

In addressing hadronic interactions in nuclear and astrophysical systems, a true Cheshire Cat Mechanism posits that ``the bosonic and fermionic degrees of freedom  can  in no way be independently constrained at the boundary"~\cite{TCCM}. {\it This would  imply that the boundary terms should be {\it necessarily} determined by locality, renormalizability and hermiticity, plus appropriate symmetries of QCD}. This means that the bag wall, therefore the confinement radius, cannot be a physically relevant quantity. 

The principal idea for what the Cheshire Cat  is all about can be glimpsed in an amusing pedestrian presentation given in 1987 Les Houches  meeting~\cite{nielsen-wirzba} where it is stated that   ``The Cheshire Cat point of view is to postulate that the bag itself of the hybrid bag model does not exist, in analogy to the non-existence of the Cheshire Cat in the Lewis Caroll fable. Only the grin of the Cheshire Cat and the formalism of the bag exist~\cite{CCP}."  The application of the idea to nuclear physics up to year $\sim 2000$ is partly summarized in \cite{MRCCP}. Here I will recount some of what has transpired since then, going into the regime of nuclear physics that has not yet been explored.
%
%%%%%doneuptohere %%%%%

\subsection{The chiral bag}
The gist of the basic point can be illustrated with  a (1+1)-dimensional CC involving an ``infinite hotel." . 

The chiral bag action consists of three components 
\be
S=S_{V}+S_{\bar{V}}+S_{\partial V}\label{S}
\ee
where the first term is of the bag of volume $V$ containing $N_c=3$  massless quarks $\psi$ for the baryon, the second is the outside of the bag of volume $\bar{V}$ populated by mesons (massless pions $\pi$ -- for now,  later heavier mesons will be brought in) and the third is the boundary connecting the inside to the outside. 

What enters in the first and second terms of the  action is known classically.  What crucially matters is the boundary term which makes the Cheshire Cat Mechanism (CCM) operative. As it can be seen easiest for the issue, we proceed in 1 space dimension focussing on a single quark but it applies to all $N_c$ of them.  The argument given for (1+1) dimensions applies equally to (1+3)d for the flavor number $ \geq  2$. The subtlety for $N_f=1$, a totally new element, will be postponed to later.
%%%%%%%

With the appropriate constraints, e.g.,    chiral symmetries etc.,  imposed in the action (\ref{S}), the problem arises in the boundary term because there is a quantum anomaly on the vector current that triggers the leakage of the baryon charge confined in $V$ into $\bar{V}$. In \cite{nielsen-wirzba}, this scenario is, amusingly, depicted as a ``jailed" quark trying to escape through the jail boundary exploiting the ``infinite hotel," the anomaly.  The consistency of the theory requires the anomaly generated at the bag side be cancelled by the pion field in $\bar{V}$ wherever the bag boundary might be located and this is done by the pion field ``turning topological" and picking up the leaked charge.  This action is independent of the bag location  $R_{bag/con}$ and this renders the bag/confinement size physically irrelevant. The baryon charge, carried by the quarks confined inside the bag, disappears into the the pion field, with the latter carrying the soliton topology that leaks out. The baryon charge therefore is the smile of the Cheshire Cat. 
\subsection{The  {\it Smile} of CC is topological}
In (1+1)d, there is an exact bosonization of the fermion quark to a boson so the baryon charge $B$  is 1 independently of the confinement wall at $R_{bag}$. Neither the charge in the bag nor the charge in the pion field separately  is physical while the total is. When the bag is shrunk to zero radius, the baryon charge remains in the smile of the cat.  Given the exact bosonization in (1+1) dimensions, it is possible to see that not just the baryon charge but also certain other  conserved quantities remain independent of the  $R_{bag}$.  But this is not feasible in (1+3) dimensions.

How this phenomenon occurs is nicely illustrated in \cite{nielsen-wirzba}. Consider a massless quark $\psi$ swimming in attempt to escape from the ``jail" to the positive direction on top of the Dirac sea. It gets blocked when it reaches the wall at $R_{bag}$. Because the symmetries do  not allow it to turn and swim back on the surface,  it is forced to stay in the sea,  but the Dirac sea is completely occupied, so how does it wiggle itself out below the Dirac surface? This is where the quantum anomaly kicks in at the boundary, allowing the quark to escape exploiting  that the ``hotel" is infinite.  The fermion number is conserved, so the quark disappears into the ``smile" of the CC.  What takes place there can be  exactly described thanks to the bosonization technique that exists in (1+1) dimensions. 
\subsection{How CC works in real world}
%%%%%%
Now what about in (1+3) dimensions where the exact bosonization is not known to exist?  The early efforts to address this issue in nuclear physics are reviewed in \cite{MRCCP}.

It was first seen in \cite{RGB} at the pion'a magic chiral angle $\theta=\pi/2$ that exactly half of the baryon charge $B=1/2$ leaks out of the bag which then is picked up by a half-skyrmion. This magic-angle phenomenon holding trivially -- thanks to the topology involved -- for the baryon charge in both 1 and 3 space dimensions has a very interesting phenomenon particularly at high density. We will return to it below. 

It was shown by Goldstone and Jaffe~\cite{GJ} using the multiple reflection method that  the leaking baryon charge from the bag in (1+3) dimension is {\it precisely picked up at any chiral angle} by the fractionalized skyrmion charge in the exterior of the bag so that the total baryon charge remains intact. As in the (1+1) dimensions what is in action is the anomaly. The leaking charge which of course depends on the bag radius $R_{bag}$ is precisely carried by the pion field (skyrmion) also in (1+3)d. This must happen because the baryon charge is a  topologically conserved quantity and it is the gauge invariance that tackles the anomaly that does the job. 

Now the question is: What about other physical quantities that are not {\it directly} associated with topology?
\subsection{CC Mechanism  as gauge invariance in action:\\ Flavor-singlet axial charge of the proton}
There is no exact bosonizaton known in (1+3))d, so it is not very likely that apart from what's involved with topology, one can expect the CCP to hold for {\it all} physical observables.  Surprisingly there are certain properties that are not directly controlled by topology but seem however to come rather close to having the CCP work. Let me describe one case where while topology is not evidently involved  there is an anomaly present at the boundary.  Here  a conserved charge does play a role, namely, the flavor-singlet axial coupling constant (FSAC) $g_A^{(0)}$ that involves the $U_A (1)$ anomaly.  With the modern progress toward ``first-principles" approaches to nuclear structure, calculations could be done a lot more accurately but even at this early stage of development with the technique which is more or less crude it becomes surprisngly clear that intricate details, not obviously topologically connected, need to be as fully treated as possible to see how the (albeit approximate) CCP can be arrived at. The FSAC, described below, gives an excellent illustration of this case.
\subsubsection{\it The $\eta^\prime$ from the $U_A(1)$ anomaly}\label{etaprime}
One prominent illustration of what could be involved in the working of the CCP is the case of the flavor-singlet axial charge $g_A^{(0)}$ of the proton measured in the EMC at CERN~\cite{protonspin} -- erroneously, as it turned out, linked to the proton spin crisis~\cite{EL}.  It is an extremely interesting case where the leakage of color from the bag and the flavor-singlet $\eta^\prime$ excitation outside can be described by a Cheshire Cat mechanism involving the $U_A(1)$ anomaly~\cite{cheshireeta}.

Let me first give a simple description of the $\eta^\prime$ properties combining the $U_A(1)$ anomaly with the CCP.  It illustrates  what the CCP states at low energy: ``hadronic processes do not discriminate between QCD degrees of freedom (quarks, gluons) on one hand and meson degrees of freedom (pions, vector mesons etc.) on the other, provided all the necessary quantum effects are properly taken into account~\cite{TCCM}." The basic idea is to consider a slowly-fluctuating  $\eta^\prime$ configuration in the vacuum. Over a small volume of size $V$,  applying the CCP allows one then to describe $\eta^\prime$ excitation either in  terms of QCD variables or in terms of effective bosonized degrees of freedom. As in the case of the baryon charge, the presence of $\eta^\prime$ at the boundary makes the color charge of the quarks to leak-out of $V$~\cite{colorleakage}. This symmetry-breaking term which involves the $\eta^\prime$ coupling to the Chern-Simons current $K_5^\mu$ 
\be
K_5^\mu=\epsilon^{\mu\nu\alpha\beta} (A_\nu^a G^a_{\alpha\beta}-\frac 23 f^{abc}gA_\nu^a A_\alpha^bA_\beta^c), 
\ee 
where $A_\mu^a$ is the gluon field, 
needs to be cancelled by quantum effects. The boundary term reproduces the Witten-Veneziano relation~\cite{Witten,Veneziano} relating the  $\eta^\prime$ mass to the topological susceptibility $\chi$~\cite{cheshireeta}.  

That the properties of the $\eta^\prime$ emerge correctly from the CC is an important point for the $g_A^{(0)}$ treated below.  
\subsubsection{{\it A chiral bag with  $\eta^\prime$}}
We are now equipped to address the FSAC $g_A^{(0)}$ problem~\cite{FSAC}. The chiral bag model we will treat is given pictorially in Fig.~{\ref{model}.  
\begin{figure}[htb]
\vskip 0cm
\begin{center}
\includegraphics[width=12cm,angle=0]{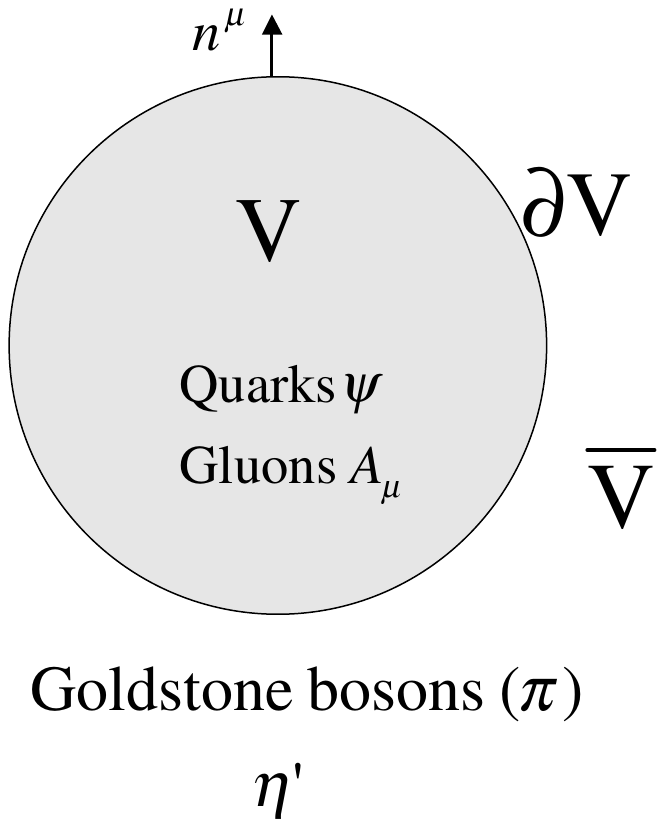} 
% \centerline{\epsfig{file=fig1.ps,width=9cm,angle=-90}}
\vskip -1.5cm \caption{\small A spherical chiral bag for
``deriving" a four-dimensional Cheshire Cat model. $\psi$
represents the doublet quark field of u and d quarks for $SU(2)$
flavor or the triplet of u, d and s for $SU(3)$ flavor and $\pi$
the triplet $\pi^+, \pi^-, \pi^0$ for $SU(2)$ and the octet
pseudoscalars $\pi$, $K$, $\bar{K}$ and $\eta$ for $SU(3)$. The
axial singlet meson $\eta^\prime$ figures in the axial
anomaly.}\label{model}
\end{center}
\end{figure}
The action contains three components as in (\ref{S}), now  in (1+3)d  
\be
S=S_{V}+S_{\bar{V}}+S_{\partial V}.\label{SS}
\ee
The first term is the usual QCD action. The other two terms are
\be
%S&=& S_V+S_{\tilde{V}}+S_{\del V},\label{cheshire}\\
%S_V&=& \int_V d^4x \left(\bar{\psi}i\not\!\!{D}\psi -\frac{1}{2}
%{\rm tr}\ F_{\mu\nu}F^{\mu\nu}\right)+\cdots\nonumber\\
S_{\tilde{V}}&=&\frac{f^2}{4}\int_{\tilde{V}} d^4x\left({\Tr}\
\del_\mu U^\dagger \del^\mu U +\frac{1}{4N_f}
m^2_{\eta^\prime}({\rm ln}U-{\rm ln}U^\dagger)^2
\right)+\cdots + S_{WZW},\nonumber\\
S_{\del V}&=&
\frac{1}{2}\int_{\del V} d\Sigma^\mu\left\{(n_\mu \bar{\psi} U^{\gamma_5}\psi)
+\frac{i g^2}{16\pi^2}{K_5}_\mu
{\Tr}\ ({\rm ln} U^\dagger-{\rm ln} U)\right\}\nonumber
\ee
where $N_f=3$ is the number of flavors,  $S_{WZW}$ is the Wess-Zumino-Witten term for $N_f\geq 3$, the ellipsis $\cdots$ stands for higher derivative terms, and
\be
U&=&e^{i\eta^\prime/f_0}e^{2i\pi/f},\\
f_0&\equiv& \sqrt{N_f/2} f. \nonumber
\ee
Here $\pi$ stands for the octet pseudo-scalar mesons and $\eta^\prime$ for the singlet pseudoscalar.
It should be mentioned at this point in anticipation of what's to come later that hidden local symmetry bosons ($\rho,\omega$), in particular, via the homogeneous Wess-Zumino term~\cite{HLS},  will figure importantly at high density.
%\footnote{It will be seen that unlike the pseudo-scalar octet mesons $\pi$ which give rise to all the octet baryons as skyrmions, the flavor singlet $\eta^\prime$ does not give a flavor singlet skyrmion baryon ${\cal B}^{(0)}$. It will lead to a fractional quantum Hall droplet which will figure at high density.} 
In the present problem, they are not important.
\begin{figure}[h]
%\vskip 2ex
\centering
\includegraphics[width=12cm]{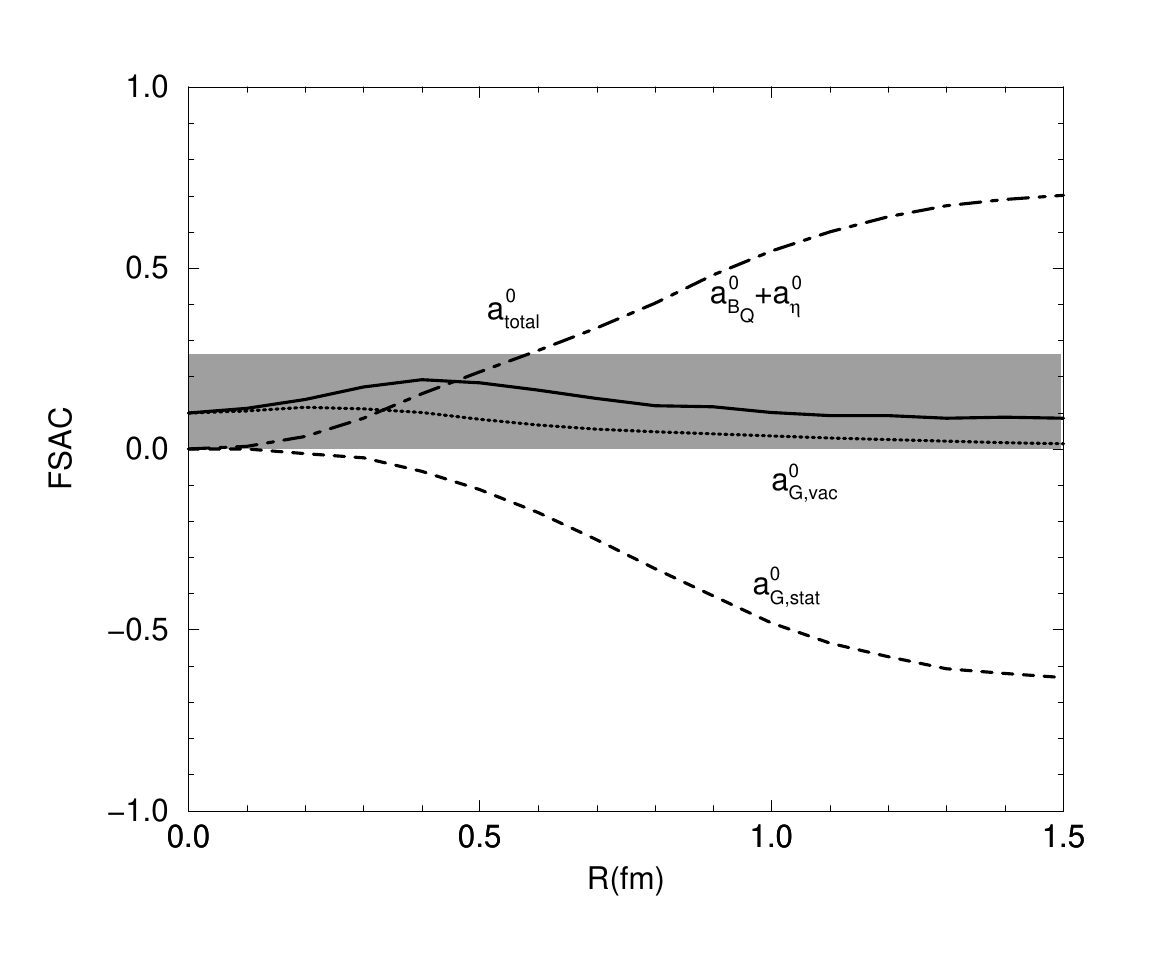}
\vskip -0.5cm
\caption{Various contributions to the flavor singlet axial constant $g_A^{(0)}$
of the proton as a function of bag radius and comparison with the experiment: (a) quark plus $\eta^\prime$ contribution ($a^0_{B_Q} +a^0_\eta$), (b) the contribution of the static gluons due to quarksource ($a^0_{G,stat}$), (c) the gluon vacuum contribution($a^0_{G,vac}$), and (d) their sum ($a^0_{total})$. The shaded area corresponds to the range admitted by experiments.}\label{picture}
\end{figure}

The matrix element of the FSAC gotten from the model (\ref{SS}) consists of terms from the bagged degrees freedom and the $\eta^\prime$ satisfying the $U_A(1)$ anomaly. As detailed in \cite{FSAC}, there are (1) the contribution from the bagged quark, $a^0_{Bq}$, with the baryon charge leakage properly taken into account, (2) the $\eta^\prime$ contribution $a^0_\eta$, (3) the contribution of static gluons due to quark source $a^0_{G,stat}$, and (4) the gluon Casimir $a^0_{G,vac}$.  The results obtained (with available parameters) for bag radius $0 < R < 1.5$ fm are plotted together with  the sum $a^0_{total}$ in Fig.~\ref{picture}. The shaded area corresponds to the EMC result for $g_A^{(0)}$.
It shows  that  the CC-ism is extremely sensitive to the detailed account of various contributions. It involves significant cancellations for the  CCP to to work out. As far as I am aware this is the only calculation in the literature that shows how the CC feature is captured numerically for quantities that are not purely topological. It would be theoretically challenging and meaningful to repeat this calculation in a more rigorous treatment. 

\subsection{CC Mechanism for finite nuclei and nuclear matter}
The CCP states that when the bag radius is shrunk to zero, the CC smile remains and physics should not depend on the ``confinement radius" in the sense implied in the MIT model. Therefore nuclei should be describable in terms of the  smiles of the multiple CCs with the bags shrunk to points. Let me remark only briefly on this issue which deserves a lot more space than available.

Nuclear physics in terms entirely of skyrmions is currently becoming a new frontier in the field. It could be considered as  of a ``first-principles" approach in nuclear physics on par with $\chi$EFT in QCD at low energy~\cite{vankolck}  in the sense of Weinberg's ``Folk Theorem" of QCD~\cite{WeinbergEFT}. There is a great deal of developments going on  and expected to come. How this picture works has been extensively studied with the Skyrme model with only the quartic derivative term (referred in the literature as Skyrme term) with some remarkable success in various observables~\cite{battye,manton}.  It is now well established that the Skyrme quartic term alone would fail to give nuclear binding energies correctly. It gives too big a binding energy and some of the ground-state properties rather poorly. This defect can be cured to a certain extent by the role of heavy vector mesons incorporated into the mesonic sector~\cite{sutcliffe} and holographic dual models with infinite towers of vector mesons~\cite{HRYY,SS}. But as will become clearer in what follows, there is a great deal of what other degrees of freedom -- and with what subtleties -- need to be figured out.

The early effort to access dense matter, particularly compact star matter, is found in \cite{park-vento}.  Only very recently are some challenging efforts  made in mathematical rigor in the skyrmion approach to  dense matter in compact stars etc. starting  to appear in the literature~\cite{adametal}.  For instance, conformal symmetry found to play a significant role in the CCP -- and treated further below -- is yet to be incorporated in these efforts and would certainly impact on some of the topics covered below. An example of what might be in store is given below. 
\subsubsection{\it Precocious pseudo-conformailty}
There is a precocious onset of conformality which begs to be clarified and is presumably connected to the CC smiles at high density.   It could be a macroscopic rendition of what's taking place in the microscopically fluctuating behavior as one encounters in the sound speed in the interior of compact stars appearing above the skyrmion-half-skyrmion  transition at $n\sim n_{1/2}$.   This is observed in a crystal-lattice simulation of the skyrmion matter above the density $n_{1/2}$ which takes place roughly at $n_{1/2} \gsim 3 n_0$ where skyrmions transform to half-skyrmions. It is one of the phenomena in skyrmion matter that remains quite mysterious and intriguing and not yet understood, but it is believed to be one of the essential features in probing topology in CCM. Let me make here a comment,  somewhat detailed, on this matter. 
 
The half-skyrmion phase in the skyrmion-crystal simulation of dense baryonic matter is, it seems most likely,  in a state that can be described {\it reliably} by mean fields,  largely undistorted by strong nuclear interactions. This phenomenon, discussed in detail already in \cite{PKLMR}, and given a brief reminder below in Sect.~\ref{GnEFT}, is found in Landau-Fermi liquid fixed point theory where the $\beta$ function for the quasiparticle interactions is suppressed. This striking feature was first found in the Skyrme model  with the Atiyah-Manton ansatz in \cite{atiyah-manton-skyrmion}.  Let me illustrate the result from \cite{PKLMR}.

Write the chiral field $U$ as $U(\vec{x}) = \phi_0(x,\, y,\, z ) + i \phi^j_\pi(x,\, y,\, z )\tau^j$  with the Pauli matrix $\tau^j$ and $j=1,2,3$. Including $\rho$ and $\omega$, we write the fields placed in the lattice size $L$ as $\phi_{\eta,\, L}(\vec{x}\,)$ with $\eta =0,\, \pi,\, \rho,\, \omega$ and normalize them with respect to their maximum values denoted $\phi_{\eta,L,{\rm max}}$ for given $L$.  It can be shown, as in \cite{atiyah-manton-skyrmion}, that in the half-skyrmion phase with $L\lsim L_{1/2}$ where  $L_{1/2}\simeq2.9$ fm, the field configurations are invariant under scaling in density as the lattice is scaled from $L_1$ to $L_2$
\be
\frac{\phi_{\eta,\,L_1}(L_1\vec{t}\,)}{\phi_{\eta,\,L_1,\,{\rm max}}}= \frac{\phi_{\eta,\,L_2}(L_2\vec{t}\,)}{\phi_{\eta,\,L_2,\,{\rm max}}}. \label{scaling}
\ee
Since other fields are quite similar, shown in  Fig.~\ref{scale_inv} is the case of $\phi_{0,\pi}$ for  $\phi_{0,\pi} (t, 0,0)$ vs. $t$ with $t\equiv x/L$.  What is seen there is that denisty-scale invariance sets in for $L\lsim L_{1/2}$. One can see that the field is independent of density in the half-skyrmion phase with $L\lsim L_{1/2}$  whereas for the skyrmion phase with lower density with $L >  L_{1/2}$, it is appreciably dependent on density.
\begin{figure}[h]
\begin{center}
\includegraphics[width=6.0cm]{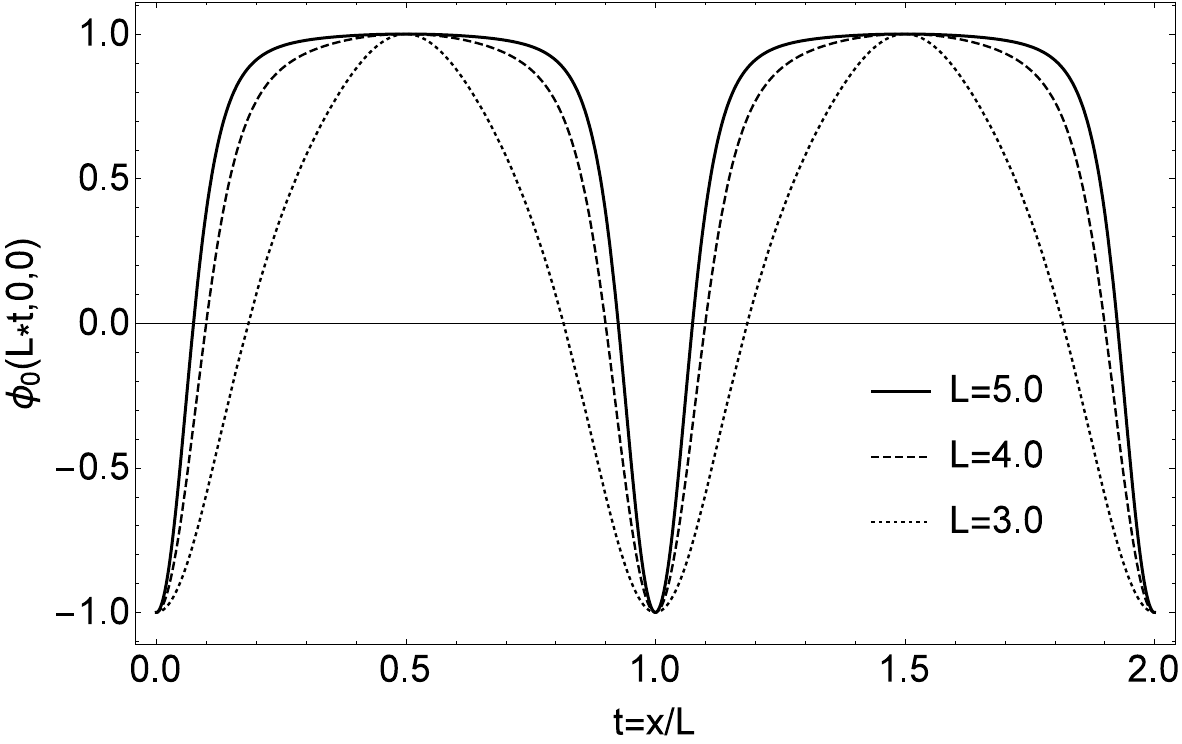}\includegraphics[width=6.0cm]{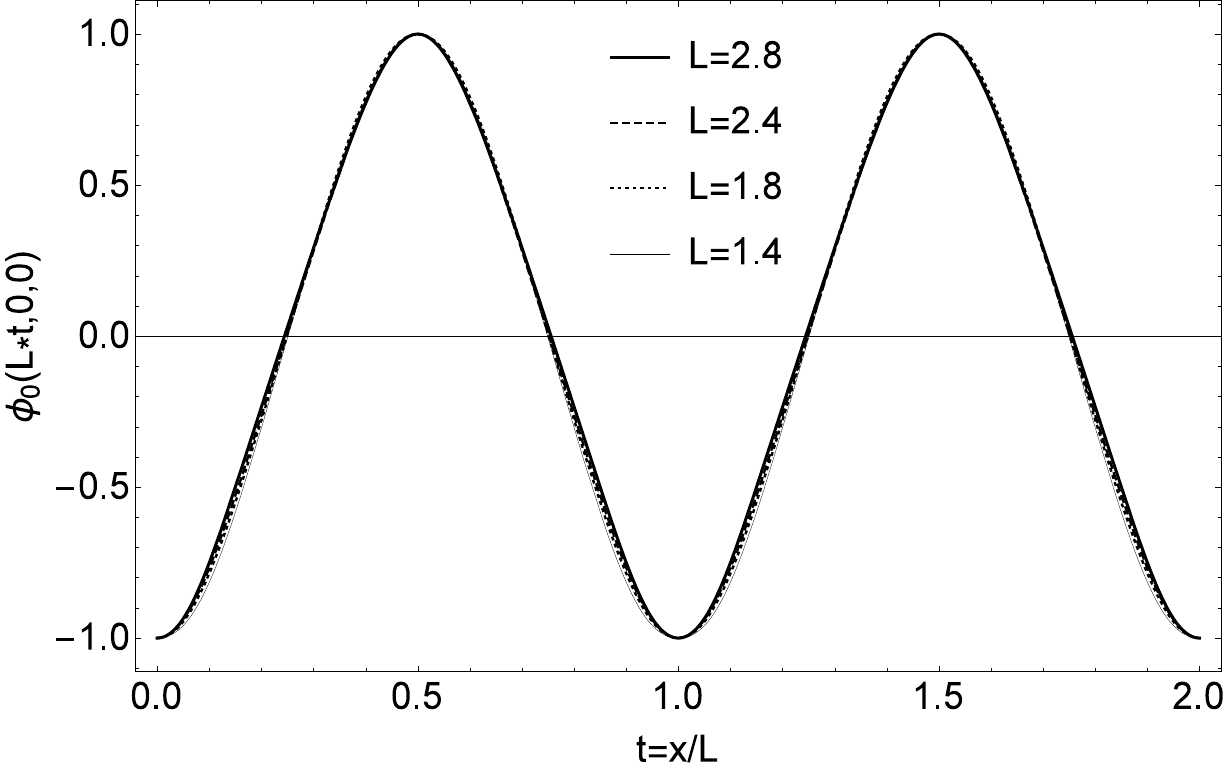}
\includegraphics[width=6.0cm]{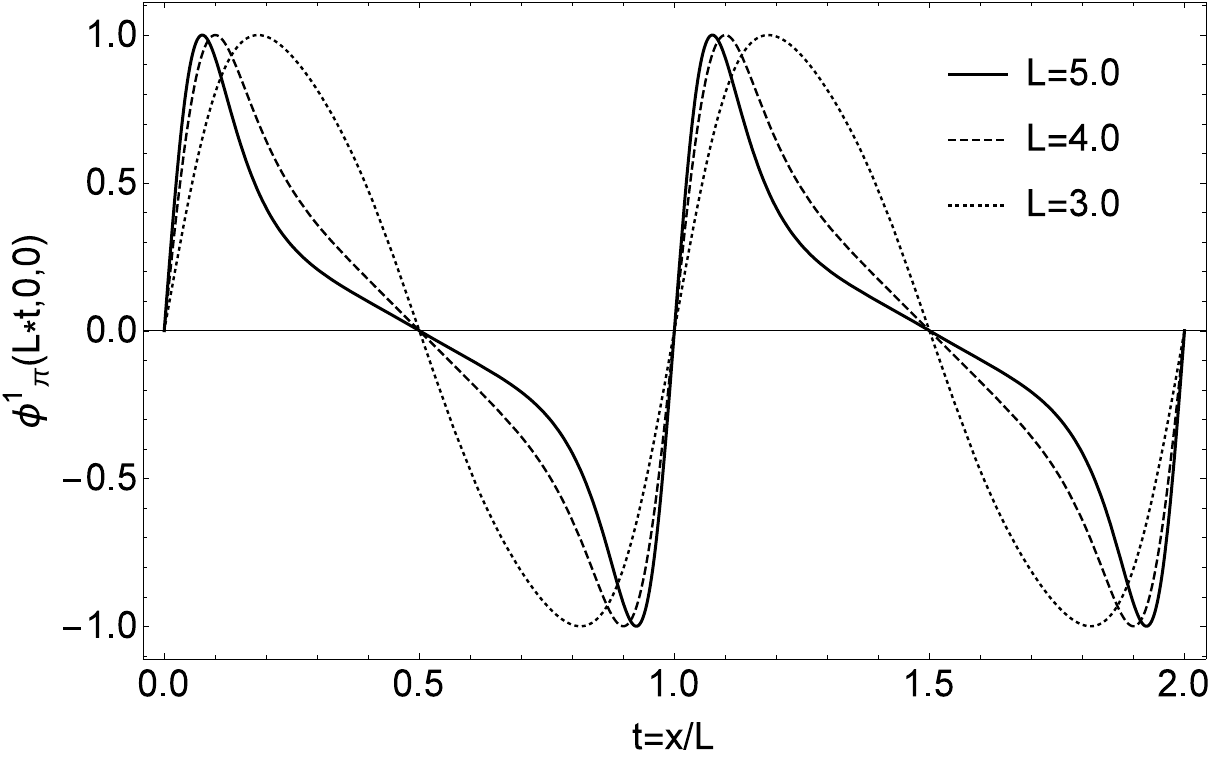}\includegraphics[width=6.0cm]{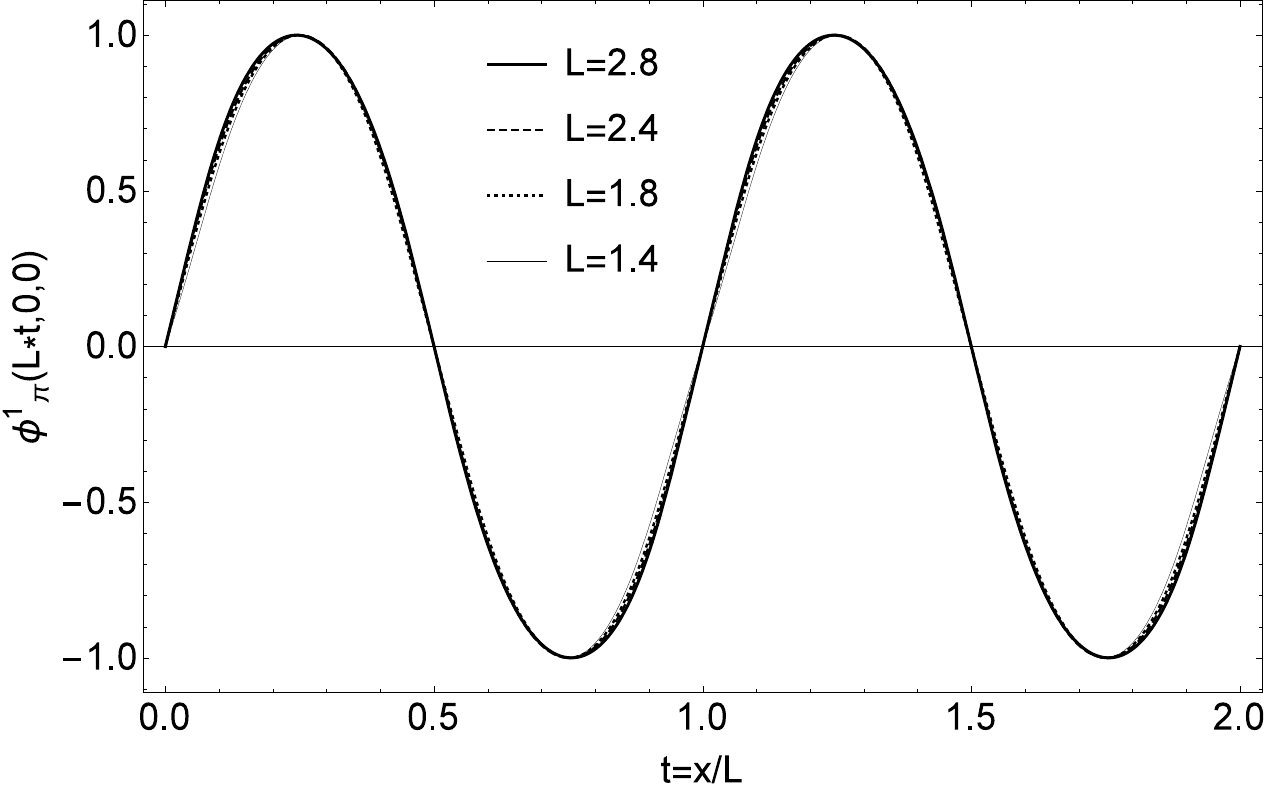}
\caption{  The field configurations $\phi_0$ and $\phi^1_\pi$  as a function of $t = x/L$ along the y = z = 0 line. The maximum values for $\eta=0,\pi$ are $\phi_{0,\,L,\, {\rm max}} = \phi_{\pi,\,L,\, {\rm max}} = 1$. The half-skyrmion phase sets in when $L=L_{1/2} \lsim 2.9\,{\rm fm}$.
 }\label{scale_inv}
 \end{center}
\end{figure}

What does this imply for the energy density?

The energy density for the skyrmion matter put on the lattice of lattice size $L$ can be written as
\begin{equation}
\epsilon = E/A/V(=L^3) = \frac{1}{L^3} \int^L_0 d^3x \sum_{n,\,m} c_{n,m}\, f_{n,m}\left( \vec{\nabla}_x,\, \phi_{\eta,\,L}(\vec{x}\,)\right)\,,
\end{equation}
where $c_{n,m}$ is the coefficient of $f_{n,m}$ which is the function of $\vec{\nabla}_x$ and $\phi_{\eta,\,L}(\vec{x}\,)$ having $n$th power of $\vec{\nabla}_x$ and $m$th power $\phi_{\eta,\,L}(\vec{x}\,)$ with $\nabla_{x,\,j} = \frac{\partial}{\partial\,x^j}$. One can reduce it to~\cite{PKLMR}
\begin{eqnarray}
\epsilon
%&=& \sum_{n,\,m} \left(\frac{1}{L} \right)^n \left(\phi_{\eta,\,L,\, {\rm max}} \right)^m \int^L_0 \frac{d^3x}{L^3}\, c_{n,m}\, f_{n,m}\left( L \vec{\nabla}_x,\, \frac{\phi_{\eta,\,L}(\vec{x}\,)}{\phi_{\eta,\,L,\, {\rm max}}}\right) \\
%&=& \sum_{n,\,m} \left(\frac{1}{L} \right)^n \left(\phi_{\eta,\,L,\, {\rm max}} \right)^m \int^1_0 d^3t\, c_{n,m}\, f_{n,m}\left( \vec{\nabla}_t,\, \frac{\phi_{\eta,\,L}(L\vec{t}\,)}{\phi_{\eta,\,L,\, {\rm max}}}\right) \\
= \sum_{n,\,m} \left(\frac{1}{L} \right)^n \left(\phi_{\eta,\,L,\, {\rm max}} \right)^m  A_{n,\,m}\,, \label{energy_den}
\end{eqnarray}
 where $A_{n,\,m}$ is a constant independent of the lattice size $L$.

 Calculating the energy density (\ref{energy_den}) in skyrmion-crystal simulations involves field configurations satisfying their equations of motion. Hence (\ref{energy_den}) is a mean field expression. It captures all essential dynamics in terms of the mean fields of each degrees of freedom involved, with residual interactions suppressed. This is in the same spirit as the Landau Fermi-liquid fixed-point  approximation that underlies G$n$EFT.  The density dependence lies, apart from the $(1/L)^n$ factor, in the maximum field configuration $\left(\phi_{\eta,\,L,\, {\rm max}} \right)^m$. This implies that in the half-skyrmion phase,  considered to set in at high density, the mean-field structure dominates. This agrees with the lore that at high density -- and in the large $N_c$ limit, the skyrmion crystal picture becomes valid  in QCD. In clear contrast, however, as one can see in Fig.~\ref{scale_inv}, the mean-filed structure breaks down in the lower-density phase with $L > L_{1/2}$. Taking the RMF approximation to be more or less equivalent to Fermi-liquid fixed point theory at some high density difficult to pin down numerically, one can take the breakdown of the mean-field structure  as perhaps a signal for non-Fermi liquid structure. 
 
In short, while it is far from quantitative and short in rigor, it is intriguing that the precocious pseudo-conformality seems to set in the half-skyrmion property (\ref{energy_den}) in G$n$EFT as  in the sound velocity in compact stars~\cite{PKLMR} and also in the long-standing puzzle of the quenched $g_A$ in nuclei~\cite{shao-rho}. Whether or not these features survive in  ``first-principles" calculations or in Nature remains to be seen. They are not yet ruled out. 

\subsection{CC as a gauge symmetry}\label{CCgauge}
 To reiterate what has been developed up to here and to be developed further,  the CCP implies that hadronic processes do not discriminate between QCD degrees of freedom (quarks, gluons) on one hand and meson degrees of freedom (pions, vector mesons etc.) on the other, provided all the necessary quantum effects are properly taken into account. 
 An appealing suggestion for what may be involved is an action of  gauge symmetry~\cite{smooth-bosonization}. A a specific case is the MIT bag radius $R_{bag}$ associated with confinement. By introducing the bag radius expressed in terms of  a collective field as  a gauge degree of freedom, a given value of $R_{bag}$ in the sense of the MIT bag with confined quarks can be taken as a gauge-fixing as in the cloudy bag~\cite{CloudyBM}. With chiral as well other symmetric couplings to appropriate degrees freedom $\pi$, vector mesons etc. without the space divided into inside and outside but suitably fitted to data, a model in the class of cloudy bag (with no boundary conditions) could potentially be formulated as a gauge-fixed CC gauge theory. Of course it would be more powerful if it could be not just for the confinement size but in a broader sense as suggested in \cite{smooth-bosonization}.  It would be interesting to see how far one can proceed in confronting Nature.
 
\section{$\eta^\prime$ and Fractional Quantum Hall Droplet}
\subsection{CC for quantum Hall  droplet}
We have seen that the flavor singlet meson $\eta^\prime$ played a crucial role for the color symmetry for baryons and  for the FSAC $g_A^{(0)}$.   In Sect.~\ref{etaprime} the properties of $\eta^\prime$ were derived  by  CCP. Now the question is why there is no skyrmion baryon coming via CC from the $\eta^\prime$ as there are  octet skyrmions coming from the octet mesons $\pi$?  There is no description of the flavor-singlet  baryon ${\cal B}^{(0)}$ in the baryon multiplet whereas the $\eta^\prime$ figures as a singlet in the multiplet of the octet $\pi$. This may be  due to the role of the color symmetry in QCD~\cite{noB0}. It is surprising how this is implied by the homotopy group  $\pi_3 (U(1))=0$ as opposed to $\pi_3(SU(2))=\pi_3 (S^3)=\cal{Z}$. 

This however does not mean that there is no topological baryonic object coming from the $\eta^\prime$.

It gas been recently pointed out by Komargodski~\cite{ZOHAR}  that for $N_f=1$  the chiral effective theory is dominated by the axial U(1) anomaly, and  the topological charge cannot be identified. It was noted that the presence of stable superselection rules in the
QCD  vacuum (instanton tunneling between vacua with different Chern-Simons number) implies  the existence of 1(time) +2 (space) dimensional domain walls. These walls connect vacua with different Chern-Simons number and are observed to be stable at large $N_c$. When these sheets are finite dimensional with a boundary, they are found to 
carry massless edge excitations with baryon quantum numbers. They  are identified with fast spinning baryons. These sheets, unconnected to skyrmions,  are  described by
a topological field theory through a level-rank duality argument, much like in the fractional quantum Hall (FQH) effect~\cite{CMHALL}. The
 baryons are analogous to the gapless edge excitations in quantum Hall (QH) droplets. 
 
 \begin{figure}[h!]
\begin{center}
 \includegraphics[width=9cm]{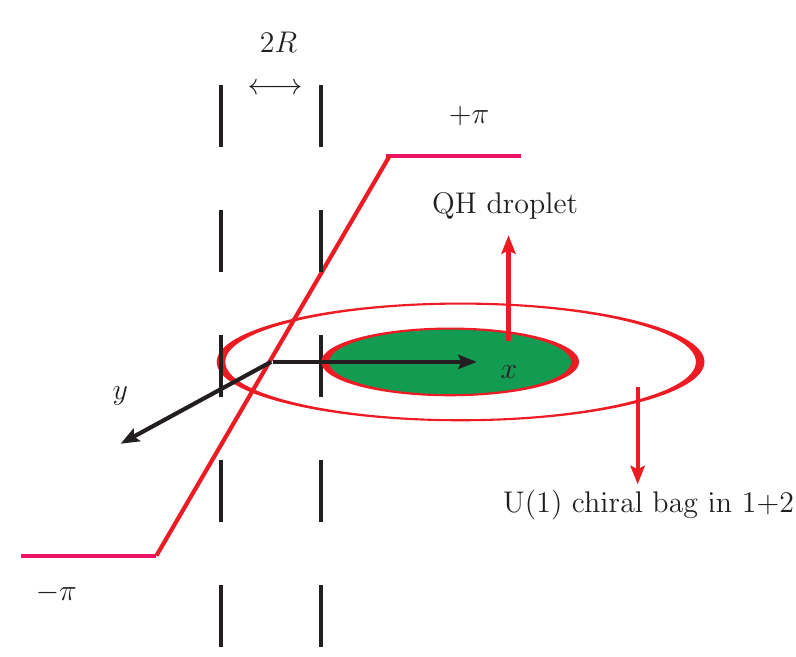}
  \caption{1+2 dimensional chiral bag surrounding a quantum Hall droplet. The bag is an annulus of width $2R$ clouded by
   an $\eta^\prime$ with a monodromy of $2\pi$. In the limit of zero bag radius, the chiral bag reduces to a vortex string with unit baryon number $b=1$.}
    \label{droplet}
 \end{center}
\end{figure}
 
 Most remarkably this quantum Hall droplet (variably called  pita or pancake) can be obtained by the CCP~\cite{MNRZ}.  A chiral bag with a single-quark species of charge $\it {b}$  (a fermion number) confined to a 1+2 dimensional annulus is found to leak the fermion charge in a way different from the skyrmion falling into an ``infinite hotel"~\cite{nielsen-wirzba}. When the bag radius is shrunk to zero,  the ``smile" of the cat  is left with spinning gapless  quarks running  luminally, accounting for the edge modes and their chirality~\cite{ZOHAR}. A current transverse to the smile is shown to appear, embodying the Callan-Harvey anomaly out-flow~\cite{CALLAN}, well-known to string theorists. This transverse current is shown to be quite analogous to the Hall current typical of the QH effect through the emergence of an effective U(1)gauge field, say, Chern-Simons field. This U(1)  gauge field lives in the disc enclosed by the Cheshire cat  smile, and is described by a purely topological field theory in 1+2 dimensions. The quantum numbers of this baryon as a QH droplet follow readily from the chiral bag construction. The droplet outside the bag is illustrated  in Fig.~\ref{droplet}. 

The QH droplet is found to be captured by the  action 
\be
S_{1+2}=\frac{{\it b}^2}{4\pi}\int_{1+2} AdA\label{Chern-Simons}
\ee
where $A_\mu$ is the abelian Chern-Simons field~\cite{CMHALL}. It is to  be noted, as elaborated later,  that this $U(1)$ field is an {\it emergent}  field dual to the $\omega$ field in HLS theory. What results for $N_f=1$ is a topological object but quite different from skyrmions for $N_f\ge 2$.  In the CC model Fig.~\ref{droplet},  the anomaly out-flow is to take place instead of the ``infinite hotel" mechanism taken for the skyrmion. The mathematical constraint $\pi_3 (U(1))=0$ deals with the singlet in the flavor space, with no other quantum numbers involved. In QCD, the color symmetry seems to figure but how it corresponds to the mathematics is unclear. That the QH droplet/pancake is not visible at low energy in nuclear processes could be because it's unstable or high-lying in energy. In QCD it is the color that could be responsible for the flavor singlet ${\cal{B}}^0$ not to join the octet skyrmion baryons coming from the octet $\pi$~\cite{noB0}.
\subsection{Tale of two hotels}
In \cite{colorleakage} a chiral bag model was constructed to prevent the color charge  from leaking from the bag following the CCP.  A boundary term was added to the chiral bag to seal  the leaking charge. This boundary term can be readily obtained by noting that for $y$-independent fields, Ref.~\cite{colorleakage}  describes the {\bf outside} of the bag as a line segment in 1+1 dimensions with
\be
 \frac {\it b}{2\pi}\int_{1+1}Ad\theta=\frac {\it b}{2\pi}\int_{1+1} \theta F-\frac {\it b}{2\pi}\int_{B}A_0\theta
\label{12X}
\ee
after an integration by parts, clearly showing the leaking of the {\it b}-charge through the boundary. To seal the leak, the {\bf inside} of the bag has to be supplemented
by the opposite boundary term
\be
\frac {\it b}{2\pi}\int_B nA\frac {\eta^\prime} {f_\eta}  \equiv -\frac {\it b}{2\pi}\int_B  \epsilon^{\mu\nu}n_\nu A_\mu\frac {\eta^\prime} {f_\eta}
\label{13X}
\ee
with $n^\nu$ the spatial normal to the bag boundary, after enforcing covariance on the 2-form.
This is exactly the surface term suggested in the Cheshire cat construction.  The present arguments illustrate the subtle relationship between the chiral bag in \cite{CCP} for the baryon charge as well as  \cite{colorleakage} for the color charge and the present chiral bag for the baryon as a FQH droplet. In the former the {\it b}-charge is absolutely confined, while in the latter the {\it b}-charge is allowed to flow transversely, both making use of a Chern-Simons term. This is the tale of two hotels: the infinite hotel in our world for the confined anomaly, and the finite hotel in the other world for the flowing anomaly. This tale is highly relevant for our formulation of nuclear and astrophysical processes involving hadron-quark continuity. For instance, the role of  the $\eta^\prime$ for the color charge conservation is responsible for the Cheshire Cat mechanism for the {\it tiny} flavor singlet axial charge for the proton $g_A^{(0)}$ described in Fig.~\ref{picture}. Furthermore since the $\eta^\prime$ is expected to become light at high density, it could have a strong impact on the stiffness of the equation of state in compact-star matter required for the observed massive $\gsim 2 M_\odot$ stars as discussed in \cite{PCMprediction}. This possibility has not  been considered up to date.

Now the question is: Are the two hotels connected and if so how?

\subsection{To go from  skyrmions to fractional quantum Hall droplets}\label{skyrmion-to-QH}
Several interesting questions are raised at this point. First can the skyrmions in the $N_f=2$ baryons, say, p and n, be made to approach the QH droplet structure  by tweaking parameters of the chiral Lagrangian? 

This issue is addressed by Karasik  in two ways~\cite{karasik}. 

First is to start with the two-flavor skyrmions in the presence of $\eta^\prime$ field with zero quark masses $m_u=m_d=0$ in the large $N_c$ limit. The mass $m_d$ is then continuously increased by fiat from 0 to $\infty$ while keeping $m_u$ at zero. The tweaking makes the theory continuously move to  a chiral compact boson living on a boundary, leading to a Chern-Simons theory. The process is quite intricate, relying on a certain amount of involved arguments and reproduces the pancake structure of \cite{ZOHAR}. It also predicts the spin of ${\cal{B}}^0$ to be $N_c/2$, the same as the decuplet $\Delta$. Of course this by itself is not a controllable operation in reality but it shows how the QH droplet is connected to the skyrmions.

The second way to see the appearance of the QH droplet is to couple the HLS vector mesons~\cite{HLS} to the pions and the $\eta^\prime$. This way, as I will describe below, offers a possibility -- the way known to the author -- to probe the QH droplet by going to high density.

Concisely stated, the key observation Karasik makes is that the cusp singularity in the potential $V_{\eta^\prime}$ in the effective Lagrangian for $\eta^\prime$ generated as gluon fields jump from one vacuum to another
\be
V_{\eta^\prime} \approx \frac 12 f_\pi^2 m_{\eta^\prime}^2  {\rm min}_{k\in Z} (\eta^\prime - 2\pi k)^2 +\cdots\label{cusp}
\ee
can be eliminated by the vector mesons in hidden local symmetry~\cite{HLS}  in the effective theory which play the role of Seiberg dual to the gluons~\cite{sdual}.  This leads to the fractional quantum Hall structure on the $\eta^\prime$ sheet as a configuration carrying a baryon charge that interpolates between $\eta^\prime=0$ and $\eta^\prime=2\pi$ captured by a $U(1)_{N_c}$ Chern-Simons theory~\cite{ZOHAR}. This structure is also indicated in holographic dual QCD~\cite{dualqcd}. 

The power of exploiting hidden local symmetry (HLS) and hidden scale symmetry (HSS) together with the notion -- up to date lacking rigorous formulation -- of vector dominance (VD) is that there is a possibility of probing the QH droplet, more precisely, fractional QH sheet structure via an effective field theory suitably applicable to dense nuclear matter relevant to massive compact stars.  What figures is the homogenous Wess-Zumino term gauge-invariant with the global symmetries and with time reversal symmetry that acts on the field $U$, $U\leftrightarrow  U^\dagger$.  Limited to $N_f=2$ for convenience, the 5-d Wess-Zumino term does not enter.  Remarkably it is the $\omega$ field in HLS that will play the role of the Chern-Simons field $A_\mu$ in (\ref{Chern-Simons}).

To discuss how this can be achieved, we need a brief discussion of the effective field theory generalized from Weinberg's $\chi$EFT~\cite{WeinbergEFT,vankolck} to apply to density $n\gsim 3n_0$ relevant to compact-star matter. We called it G$n$EFT and was mentioned above..
\subsubsection{\it Generalized nuclear EFT (G$n$EFT)}\label{GnEFT}
It is now well established that the $\chi$EFT anchored on Weinberg's ``Folk Theorem" -- call it standard (S)$\chi$EFT -- works quantitatively well up to baryon density $\lsim 2n_0$ where $n_0\approx 0.16$ fm$^{-2}$  is the nuclear equilibrium density. The central density of massive compact-star matter measured in astrophysical observations could reach $\sim (5-7)n_0$. To access this density regime, however, degrees of freedom higher in scale than what figures in  S$\chi$EFT need to be introduced. Holographic dual approaches incorporate infinite towers of vector mesons. In G$n$EFT, the next most relevant degrees of freedom  taken into account are the HLS vectors $V_\mu=(\rho_\mu,\omega_\mu)$ and the hidden scale symmetry dilaton $D$ that we will represent by the ``conformal compensator field" $\chi=f_\chi e^D/f_\chi$. To this G$n$EFT,   is added the flavor singlet $\eta^\prime$.\footnote{The mixing between the octet scalar $\eta$ and the flavor singlet $\eta^\prime$ will be ignored.} What figures importantly in this approach is the degree of freedom associated with the spontaneous breaking of scale symmetry hidden in QCD. It is not usually considered explicitly but appears in S$\chi$EFT at some low order in chiral perturbation theory.

There is a long history -- and with controversy -- on the scalar dilaton resulting from the broken scale symmetry that we will choose to avoid to get into details. For our problem, we  adopt the most recent approach called ``Genuine Dilaton (GD)" by Crewther (with Tunstall)~\cite{Crewther} and/or ``QCD-Conformal Dilaton (QCD-CD)" by Zwicky~\cite{Zwicky}, both of which are anchored on assuming an IR fixed-point in QCD with $N_f\lsim 3$.\footnote{The controversy has to do with whether or not there can be an IR fixed point for $N_f \ll 8$ in QCD, away from the conformal window, being discussed re:  dilatonic Higgs in particle theory. Some current developments are at odds with the long-standing controversy. The work discussed in this article is in favor of the recent  proposals~\cite{Crewther,Zwicky} as a possibility for an emergence phenomenon in nuclear correlations.}   Applied to our case where scale symmetry is most likely  ``emergent" by strong baryonic correlations,  these two approaches are essentially the same except for certain fine details we are not concerned with. The IR fixed point we are concerned with is characterized by the presence of the Nambu-Goldstone bosons $\pi$ and $\chi$ with  nonzero decay constants, $f_\pi\neq 0$ and $f_\chi\neq 0$, thereby accommodating massive matter fields such as baryons $\psi$, vector mesons $V$ etc. The G$n$EFT was formulated to apply, at least semi-quantitatively,  from normal nuclear matter to compact-star matter. 

The effective Lagrangian that leads to the G$n$EFT will be denoted ${\cal L}_{\psi, HS,\chi}$  with HS standing for hidden symmetries, both local and scale,  and  with the $\eta^\prime$ coupled in to account for the $U_A(1)$ anomaly, it will be written ${\cal L}_{\psi, HS,\chi,\eta^\prime}$.  Now the key ingredient among the quantities that enter into ${\cal L}_{\psi, HS,\chi}$ is topological as befits the CCP. It is the topology change from skyrmions to half-skyrmions assumed to take place at the density $n_{1/2}\sim (2-3) n_0$. It encodes a wide-range of density-dependent parameters in the Lagrangian connected logically to the topological structure consistent with the CCP~\cite{MRreview}.  This many-nucleon approach is to encompass the  S$\chi$EFT and go up in density scale to the compact-star regime by implementing Landau Fermi-liquid  fixed point approximation in a generalized density functional (GDF) approach \`a la Hohenberg-Kohn theorem. In our notation, this is referred to as pseudo-conformal model, concisely  summarized  in \cite{PCMprediction}.  There one finds that its prediction, with essentially no free parameters, encounters, so far, no serious inconsistency with observables from normal nuclear matter densities to the densities measured in the recent gravity-wave and associated data in compact-star matter\footnote{Just for comparison, see the most up-to-date analyses~\cite{lattimer}.}.
\subsubsection{\it From chiral EFT to Kohn-Sham density functional \\ and the dilaton-limit fixed-point}
It may be somewhat nontrivial to those unfamiliar to how a generalized density functional can be gotten from  $\chi$EFT with hidden symmetries minimally implemented. This follows from the work of Ref. \cite{FR} based on the strategy given in \cite{shankar} on the formulation of renormalization-group approach to many-fermions on the Fermi surface.
In fact one arrives at it by a simple formulation in terms of the Landau Fermi-liquid fixed-point interactions given by BR scaling~\cite{FR}. This GDF formulation anchored on Hohenberg-Kohn theorem invoked above  is most likely closely connected to what is discussed in the fractional Quantum Hall phenomenon in condensed matter physics.   The Kohn-Sham density functional theory, widely applied in nuclear physics, turns out to have an impact in the fractional quantum Hall effect studied in mesoscopic devices as shown in \cite{hu-jain}. It would be interesting to further elucidate the possible conceptual link between the two fields.  

In approaching the FQH droplet  ${\cal B}^0$,  ${\cal L}_{\psi, HS,\chi,\eta^\prime}$ is treated in the mean-field which corresponds to the  LFL fixed-point approximation. Next    
one redefines the field $\Sigma=\frac{f_\pi}{f_\chi} U\chi$ with $U=e^{i\pi/f_\pi}$  and take the limit ${\rm Tr} (\Sigma^\dagger \Sigma)\to 0$. In order to avoid the singularities that  develope  in the limiting process, one is required to put what are referred to as the ``dilaton-limit fixed-point (DLFP)" constraints~\cite{dlfp-Bira}\footnote{In this reference the notion of DLFP was first formulated in terms of quark model but it has been verified to equally apply to quasi-nucleons as in G$n$EFT.}
\be
&& f^\ast_\pi \to  f^\ast_\chi,  \ g^\ast_{\rho NN} = g^*_{\rm COR} g^*_\rho\to 0, g^*_{\rm COR}/g^*_\rho\to 0, \nonumber\\
&& g^\ast_A\to 1, m^\ast_N\to m_0\propto f_\chi^\ast\label{DLFP}
\ee
where $g^\ast_\rho$ is the HLS gauge coupling in medium in RG analysis~\cite{HLS} and  $g^\ast_{\rm COR}$ stands for standard (many-body) nuclear correlation effects in dense matter. The $\ast$ stands for density dependence encoded in the condensates of the vacuum modified by the medium. 

There are several remarkably powerful implications in the constraints (\ref{DLFP}), some of which, it should be mentioned, have not yet been  given solid theoretical support. The first is that the DLFT indicates an approach to the IR fixed point \`a la  GD (and CD-QCD) scenario that our approach espouses. Indeed $f^\ast_\pi\to f^\ast_\chi$ deviates strongly from the conformal window scenario (relevant to dilaton Higgs model)  $f_\pi^2/f_d^2  \ll 1$~\cite{appelquist}. What's directly relevant to our approach to the FQH droplet is the decoupling of the iso-vector vector mesons in HLS  {\it before} reaching the vector manifestation fixed point $g^\ast_\rho\to 0$ at which the $\rho$ becomes the genuine gauge field.  Furthermore $g^\ast_A\to 1$ is linked to the long-standing quenching of $g_A$ in nuclear matter~\cite{gA*} and the relation $m_N^\ast\approx f^\ast_\chi$ follows from the emergence of parity-doubling symmetry in dense medium responsible for  the sound speed in the core of compact stars approaching precociously the pseudo-conformal sound velocity $v^2_{pcs}/c^2\approx 1/3$ in the core of compact stars~\cite{PCMprediction}.
\subsubsection{\it The role of ``homogeneous"/``hidden" Wess-Zumino term} 
 Now when the HLS fields are ``integrated in," apart from the 5-d Wess-Zumino-Witten term which is absent for $N_f=2$ that we are concerned with, there is a 4-d action that has the homogeneous solution to the  Wess-Zumino anomaly equation, called homogeneous\footnote{Called ``hidden" by Karasik when combined with the VD.} WZ term~\cite{HLS}. This action, conserving parity and charge conjugation but violating intrinsic parity, contains four terms including an external field constructed with Maurer-Cartan 1-forms. 
 
 What's most significant for the issue concerned here is that the vector mesons in HLS are -- most likely -- Seiberg-dual gluons~\cite{sdual} and {\it in the DLFP,  the iso-vector mesons decouple before reaching the IR fixed point} as seen in (\ref{DLFP}). That leaves the $\omega$ meson. As argued by Karasik~\cite{karasik} in terms of the domain wall with $\lim_{z\to -\infty} \eta^\prime (z)=0$ and  $ \lim_{z\to +\infty} \eta^\prime (z)=2\pi$, one  then arrives at the $(1+2)$ domain theory
\be
{\cal L}_{\rm domain-wall}= \frac{N_c}{4\pi} \omega d\omega
\ee
which is lodged in the bulk of the sheet. This reproduces the CC result (\ref{Chern-Simons}) relating the $\omega$ field to the Chern-Simons field $A_\mu$. The  VD (vector dominance) emerges in the theory fixing the arbitrary parameters in the hWZ of HLS~\cite{HLS} to hidden(h) WZ~\cite{karasik} .

What we have here is just one sheet. There is an ongoing work of sheet structure in  dense matter in the field of skyrmion crystals. There are mathematical treatments of multi-sheet structure actively studied in the field~\cite{canfora} and it is much too early to make contact with Nature. In fact such a sheet structure is already seen in skyrmion-crystal structure at finite density~\cite{paeng-park-vento} and speculated in compact-star physics where fractionally charged quasi-baryons play the role of deconfined quarks in the interior of compact stars~\cite{MR-fractional}.

In the DLFP limit, the $\rho$ mesons decouple as stressed above. In Nature, the DLFP density can be  pinned down neither by theory nor by experiments. It is likely at a lower density than that of the vector manifestation. The question then arises what happens if the iso-vector mesons are not sufficiently decoupled. It seems plausible that one would then encounter non-abelian Chern-Simons droplets before reaching the abelian Chern-Simons droplet. In fact this is what one finds in the Cheshire Cat description~\cite{MNRZ}. 
For arbitrary $N_f$ the spin and statistics arguments do not change as they are solely fixed by $N_c$. However, the leaking flavor currents lead to a U($N_f$) flavor-valued emergent gauge field ${\cal{A}_\mu}$. The CCP applies {\it mutatis mutandis}. 
The  emergent non-Abelian Chern-Simons action  would then be
\be
\frac {N_c}{4\pi}\int_{1+2} {\rm Tr}\left({\cal A } d{\cal A}+\frac 23 {\cal A}^3\right).
\label{13}
\ee
This suggests a wide range of stacked sheets of Chern-Simons droplets (or pitas) before reaching the truly deconfined quarks which may not survive the gravitational collapse. As in condensed matter systems, the phase structure as viewed in this scenario could be much richer than what can be seen top-down in perturbative QCD. It could be that ultimately which way is superior will depend on which is more efficient in capturing the vast variety of physics.
\section{Further Comments and Open Issues}
Even at the far-from-rigorous level that was adopted in this article, there are several issues that beg to be further elaborated. Let me mention a few of them here for possible future developments. In the present status of poor understanding of highly dense matter, it would be premature to rule out any of  them.

\subsubsection{\it Multifaceted role of the $\omega$ meson in nuclear matter}

In the vacuum the $\omega$ meson  is with the $\rho$ meson in hidden local symmetry. But in nuclei, it starts deviating from $U(3)$ in controlling the nuclear forces while the $\rho$ meson decouples from the dense matter before the vector manifestation fixed-point~\cite{HLS} is arrived at. It provides the crucial repulsion to counter the attraction due to the scalar (dilaton)-meson exchange as the density approaches the dilaton-limit-fixed point. It gets transformed via Vector Dominance to the abelian Chern-Simons field. How this feature gets manifested in terms of QCD variables would be important for the physics of compact-star matter. This is an open issue that has not been addressed up to date.

\subsubsection{\it Half-skymion matter}

The half-skyrmion structure figures crucially in the framework of G$n$EFT in controlling the baryonic matter at high density (as reviewed in \cite{MR-fractional}). Although it does not appear explicitly, a single skyrmion may actually be a bound two half-skyrmions by a hidden monopole as suggested in \cite{cho}. One would wonder whether  the monopole is suppressed at high density, so that one can consider deconfined half-skyrmions interacting scale-invariantly as seen in skyrmion crystal configurations in Fig.~\ref{scale_inv}. In condensed matter physics in lower dimensions, deconfined quantum critical phenomena could involve such deconfined half-skyrmions~\cite{QCP}\footnote{More recent developments do away in condensed matter physics with the deconfined half-skyrmions, going to local effective field theories directly without changing  essential physics.}. Also in CCM, the magic angle that involves half-skymions with the other half shared by the vector meson $\rho$ provides an extremely simple account to the electromagnetic form factors of the proton at high momentum transfers~\cite{magic}. It is not clear how to transform the half-skyrmions to deconfined quarks.

\subsubsection{\it Stacks of FQH droplets}

As mentioned, there are indications for the formation of the sheets of half-skyrmions in the skyrmion crystal simulations~\cite{paeng-park-vento,canfora} at increasing density. The discussion in Sect.~\ref{skyrmion-to-QH} suggested the formation of a QH droplet with $J=N_c/2$ as density is increased  toward the DLFP. It seems that there can be a stack of  QH droplets mixed with, possibly unstable, half-skyrmion sheets before reaching the DLFP. That may give rise to something like the domain-wall structure with deconfined quarks on the sheets (or droplets) and confinement in the bulk~\cite{domain-wall}.  This could lead to the precocious pseudo-conformal EoS~\cite{PCMprediction} resembling what's considered as evidence of quark-matter core in massive compact stars~\cite{evidence-core}.

\subsubsection{\it Superqualitons}

Based on that in the large $N_c$  limit, the skyrmion description for the nucleon is equivalent to the constituent quark (CQ)  description, the question was raised by Kaplan~\cite{kaplan} whether the 1/3-charged skyrmion, called qualiton, can be equated to the CQ. The answer turns out to be negative. The qualiton could not be stabilized. Then one asked whether the qualiton could not be made stable at large density. This question was answered positive~\cite{superqualiton}.

The intricate workings of skyrmions and QH sheets discussed above hinted at the possibility that the inhomogeneous structure of domain-walls, sheets etc., evolve into a homogeneous matter at super-high density. Indeed at asymptotic density, as generally accepted in particle/nuclear physics community, the color-flavor locking (CFL) must set in~\cite{CFL}, and there are a large number of papers in the literature suggesting that the CFL phase could be relevant in the properties of massive compact stars.  Regardless of whether or not the CFL is actually relevant in the observation of recent gravity-wave signals such as in \cite{evidence-core}, it was an interesting question to ask whether the CC mechanism would not tend to equate at high density the qualiton, called superqualiton, to the CQ. Indeed this has been shown to be the case~\cite{superqualiton}.  Just as pseudo-conformality seems to set in precociously in the interior of compact stars, the precocious appearance of superqualitons, i.e., fractionalized baryons, could account for the observation of quark-matter core in \cite{evidence-core}.

\subsection*{Acknowledgments}

This article was written at the invitation of the Editors of JSPC  for which I am grateful.
\subsection{Note added to the ELSEVIER version}
\indent\indent In describing the multifarious manifestation of the smile of Cheshire Cat, I made a jump skipping the intermediate states from a dense matter to the asymptotically dense matter where the CFL phase is to appear as the superqualiton phase in CCM.  One obvious case among many others is the 2-flavor color superconductivity (2SC) in CC-ized. Things can get much more complicated and also perhaps more interesting in the way the "smile" manifests before approaching the asymptotic density as was discussed. An example is discussed in \cite{QHL} on the possibility of quantum Hall liquid structure involving not baryons but flavor-singlet vector mesons with the ``smile" of the CC becoming invisible. It would be interesting to figure out how conformal symmetry affects at increasing density toward the CFL regime.

In short it seems to open up a whole new phenomena.

\subsection*{Abbreviations used}

CC: Cheshire Cat\\
CCP(M): Cheshire Cat Principle(Mechanism)\\
CQ: constituent quark\\
DLFP: dilaton-limit ffixed-point\\
(F)QH: (fractional) quantum Hall\\
FSAC: flavor singlet axial charge\\
GD: genuine dilaton\\
GDF: general density functional\\
G$n$EFT: generalized nuclear effective field theory\\
HLS: hidden local  symmetry\\
HSS: hidden scale symmetry\\
L(B)B: Little(big) Bag\\
h(h')WZ: homogeneous (hidden) Wess-Zumino


\begin{thebibliography}{42}

\bi{MITbag} K.~Johnson,
``The M.I.T. bag model,''
Acta Phys. Polon. B \textbf{6} (1975), 865
MIT-CTP-494.

\bi{LB} G.~E.~Brown and M.~Rho,
``The Little Bag,'' Phys. Lett. B \textbf{82} , 177 (1979).

\bi{weise} L.~Brandes and W.~Weise,
``Constraints on phase transitions in neutron star matter,''
Symmetry \textbf{16} (2024) no.1, 111.

\bibitem{CBM}
G.~E.~Brown and M.~Rho,
``The Chiral Bag,''
Comments Nucl. Part. Phys. \textbf{18}  no.1, 1 (1988).


\bibitem{CBMapplied}
G.~E.~Brown, M.~Rho and V.~Vento,
``Little Bag dynamics,''
Phys. Lett. B \textbf{84}, 383-388 (1979).


\bi{CCP} S.~Nadkarni, H.~B.~Nielsen and I.~Zahed,
``Bosonization relations as bag boundary conditions,''
Nucl. Phys. B \textbf{253}, 308-322 (1985); S.~Nadkarni and I.~Zahed,
``Nonabelian Cheshire Cat bag models in (1+1)-dimensions,''
Nucl. Phys. B \textbf{263} (1986), 23-36 (1985).

\bi{TCCM} S. Nadkarni and H.B. Nielsen, ``PARTIAL BOSONIZATION: The formalism of Cheshire cat bag models," Nucl. Phys. B263, 1 (1986).


\bi{skyrme} T.~H.~R.~Skyrme,
``A nonlinear field theory,''
Proc. Roy. Soc. Lond. A \textbf{260} (1961), 127-138;  ``A unified field theory of mesons and baryons,''
Nucl. Phys. \textbf{31} (1962), 556-569.

\bi{CloudyBM} A.~W.~Thomas,
``Chiral symmetry and the bag model: A new starting point for nuclear physics,''
Adv. Nucl. Phys. \textbf{13}, 1-137 (1984).

\bi{Witten-qcd} E. Witten, ``Current algebra, baryons, and quark confinement,''
Nucl. Phys. B \textbf{223}, 433-444 (1983).

\bi{noB0} J.~Est\'evez, F.~J.~Llanes-Estrada, V.~Mart\'\i{}nez-Fern\'andez and A.~Pastor-Guti\'errez,
``Lightest flavor-singlet $qqq$ baryons as witnesses to color,''
Phys. Rev. D \textbf{102}, no.11, 114032 (2020).

\bi{nielsen-wirzba} H.~B.~Nielsen and A.~Wirzba,
``The Cheshire  Cat Principle applied to hybrid bag models," in Les Houches Workshop on Theoretical Physics, March 24-April 2, 1987 (Springer Proceedings in Physics. V 26, (1988)) 


%

\bi{MRCCP}  M.~A.~ Nowak, M.~ Rho and I.~ Zahed, {\it Chiral Nuclear Dynamics}\ (World Scientific, 1996);  
M.~Rho,
``The Cheshire Cat Hadrons Revisited,''
[arXiv:hep-ph/0206003 [hep-ph]] revision and addendum to "Cheshire Cat Hadrons," Phys. Repts. {\bf 240}, 1-141 (1994).

\bi{RGB} M.~Rho, A.~S.~Goldhaber and G.~E.~Brown,
``Topological soliton bag model for baryons,''
Phys. Rev. Lett. \textbf{51}, 747-750 (1983).


\bi{GJ} J.~Goldstone and R.~L.~Jaffe,
``The baryon number in chiral bag models,''
Phys. Rev. Lett. \textbf{51}, 1518 (1983).


\bi{protonspin} B.~W.~Filippone and X.~D.~Ji,
``The spin structure of the nucleon,''
Adv. Nucl. Phys. \textbf{26}, 1 (2001).

\bi{EL} E. Leader,
``The non-existence of the proton spin crisis,''
PoS \textbf{SPIN2018}, 123 (2018).

\bi{cheshireeta} H.~B.~Nielsen, M.~Rho, A.~Wirzba and I.~Zahed,
``The tale of the eta-prime from the cheshire cat principle,''
Phys. Lett. B \textbf{281}, 345-350 (1992).

\bi{colorleakage} H.~B.~Nielsen, M.~Rho, A.~Wirzba and I.~Zahed,
``Color anomaly in a hybrid bag model,''
Phys. Lett. B \textbf{269}, 389-393 (1991).

\bi{Witten} E.~Witten,
``Current algebra theorems for the U(1) Goldstone boson,''
Nucl. Phys. B \textbf{156}, 269-283 (1979).

\bi{Veneziano} G.~Veneziano,
``U(1) without instantons,''
Nucl. Phys. B \textbf{159}, 213-224 (1979).


\bi{FSAC} H.~J.~Lee, D.~P.~Min, B.~Y.~Park, M.~Rho and V.~Vento,
``The proton spin in the chiral bag model: Casimir contribution and Cheshire cat principle,''
Nucl. Phys. A \textbf{657}, 75-94 (1999). 

\bi{HLS} M.~Harada and K.~Yamawaki,
``Hidden local symmetry at loop: A New perspective of composite gauge boson and chiral phase transition,''
Phys. Rept. \textbf{381}, 1-233 (2003).


  \bi{vankolck} U.~van Kolck,
``Nuclear effective field theories: Reverberations of the early days,''
Few Body Syst. \textbf{62}, no.4, 85 (2021);  
  H.~W.~Hammer, S.~K\"onig and U.~van Kolck,
``Nuclear effective field theory: status and perspectives,''
Rev. Mod. Phys. \textbf{92}, no.2, 025004 (2020).

\bi{WeinbergEFT} S.~Weinberg,
``What is quantum field theory, and what did we think it is?,''
[arXiv:hep-th/9702027 [hep-th]].  



 

\bi{battye} R.~A.~Battye, N.~S.~Manton, P.~M.~Sutcliffe and S.~W.~Wood,
``Light nuclei of even mass number in the Skyrme model,''
Phys. Rev. C \textbf{80}, 034323 (2009).

\bi{manton} N.~S.~Manton,
``Skyrmions \textendash{} A Theory of Nuclei,''
World Scientific, 2022.

\bi{sutcliffe} C.~Naya and P.~Sutcliffe,
``Skyrmions in models with pions and rho mesons,''
JHEP \textbf{05}, 174 (2018).


\bi{HRYY}  D.~K.~Hong, M.~Rho, H.~U.~Yee and P.~Yi,
  ``Chiral dynamics of baryons from string theory,''
  Phys.\ Rev.\ D {\bf 76}, 061901 (2007)
 % doi:10.1103/PhysRevD.76.061901
  [hep-th/0701276 [HEP-TH]].

\bi{SS} H.~Hata, T.~Sakai, S.~Sugimoto and S.~Yamato,
  ``Baryons from instantons in holographic QCD,''
  Prog.\ Theor.\ Phys.\  {\bf 117}, 1157 (2007)
 % doi:10.1143/PTP.117.1157
  [hep-th/0701280 [HEP-TH]].
  
\bi{park-vento} B.-Y. Park and V. Vento, ``Skyrmion approach to fininte density and temperature" in {\it The Multifaceted Skyrmion}\  (World Scientific, 2017) ed. by M. Rho and I. Zahed.  

\bi{adametal} C.~Adam, C.~Naya and A.~Wereszczy\'nski,
``Carbon-12 in the generalized Skyrme model,''
[arXiv:2401.08778 [nucl-th]];
C.~Adam, A.~Garcia Martin-Caro, M.~Huidobro and A.~Wereszczynski,
``Skyrme Crystals, Nuclear Matter and Compact Stars,''
Symmetry \textbf{15}, no.4, 899 (2023); 
% M.~Huidobro, P.~Leask, C.~Naya and A.~Wereszczynski,
%``Compressibility of dense matter in the $\rho$-meson variant of the Skyrme model,'' 
%C.~Adam, C.~Naya and A.~Wereszczy\'nski,
%[arXiv:2405.20757 [hep-th]]. 
C.~Adam, A.~Garc\'\i{}a Mart\'\i{}n-Caro, M.~Huidobro, R.~V\'azquez and A.~Wereszczynski,
``Kaon condensation in skyrmion matter and compact stars,''
Phys. Rev. D \textbf{107}, no.7, 074007 (2023);
C.~Adam, A.~Garc\'\i{}a Mart\'\i{}n-Caro, M.~Huidobro, R.~V\'azquez and A.~Wereszczynski,
``A new consistent neutron star equation of state from a generalized Skyrme model,''
Phys. Lett. B \textbf{811}, 135928 (2020).
%

\bi{PKLMR} W.~G.~Paeng, T.~T.~S.~Kuo, H.~K.~Lee, Y.~L.~Ma and M.~Rho,
``Scale-invariant hidden local symmetry, topology change, and dense baryonic matter. II.,''
Phys. Rev. D \textbf{96}, no.1, 014031 (2017).


\bi{atiyah-manton-skyrmion} B.~Y.~Park, D.~P.~Min, M.~Rho and V.~Vento,
  ``Atiyah-Manton approach to skyrmion matter,''
  Nucl.\ Phys.\ A {\bf 707}, 381 (2002).
  

\bi{shao-rho} L.~Q.~Shao and M.~Rho,
``Corrections to Landau Fermi-liquid fixed-point approximation in nonlinear bosonized theory:~An application to $g_A^L$ in nuclei,''
Phys. Rev. C \textbf{110}, no.1, 015204 (2024).

\bi{smooth-bosonization} P.~H.~Damgaard, H.~B.~Nielsen and R.~Sollacher,
``Smooth bosonization: The Cheshire cat revisited,''
Nucl. Phys. B \textbf{385}, 227-250 (1992).    
  

\bibitem{ZOHAR}
  Z.~Komargodski,
  ``Baryons as quantum Hall droplets,''
  arXiv:1812.09253 [hep-th].
  %%CITATION = ARXIV:1812.09253;%%
  %1 citations counted in INSPIRE as of 27 Jun 2019

\bi{CMHALL} 
  D.~Tong,
  ``Lectures on the Quantum Hall Effect,''
  arXiv:1606.06687 [hep-th].
  
 \bi{MNRZ} Y.~L.~Ma, M.~A.~Nowak, M.~Rho and I.~Zahed,
``Baryon as a quantum Hall droplet and the quark-hadron duality,''
Phys. Rev. Lett. \textbf{123}, 172301 (2019).
  
  \bibitem{CALLAN}
  C.~G.~Callan, Jr. and J.~A.~Harvey,
  ``Anomalies and fermion zero modes on strings and domain walls,''
  Nucl.\ Phys.\ B {\bf 250}, 427 (1985).
  
  
\bi{PCMprediction} M.~Rho,
``Dense baryonic matter predicted in \textquotedblleft{}pseudo-conformal model\textquotedblright{},''
Symmetry \textbf{15}, no.6, 1271 (2023)
[arXiv:2305.04715 [nucl-th]].

  
 \bi{karasik} A.~Karasik,
``Vector dominance, one flavored baryons, and QCD domain walls from the `hidden' Wess-Zumino term,''
SciPost Phys. \textbf{10}, no.6, 138 (2021); ``From skyrmions to one-flavored baryons and beyond," Symmetry 2022, 14, 2347.

\bi{sdual} Z.~Komargodski,
``Vector mesons and an interpretation of Seiberg duality,''
JHEP \textbf{02}, 019 (2011).
 
 \bi{dualqcd} F.~Bigazzi, A.~L.~Cotrone and A.~Olzi,
``Hall droplet sheets in holographic QCD,''
JHEP \textbf{02}, 194 (2023).

\bibitem{Crewther}
R.~J.~Crewther,
``Genuine dilatons in gauge theories,''
Universe \textbf{6}, no.7, 96 (2020);  
R.~J.~Crewther and L.~C.~Tunstall,
``$\Delta I=1/2$ rule for kaon decays derived from QCD infrared fixed point,''
Phys. Rev. D \textbf{91},  034016 (2015).

\bibitem{Zwicky}
R.~Zwicky,
``QCD with an infrared fixed point and a dilaton,''  Phys. Rev. D \textbf{110}, no.1, 014048 (2024)
[arXiv:2312.13761 [hep-ph]].
%L.~Del Debbio and R.~Zwicky,
%``Dilaton and massive hadrons in a conformal phase,''
%JHEP \textbf{08}, 007 (2022).

\bi{MRreview}  Y.~L.~Ma and M.~Rho,
``Towards the hadron\textendash{}quark continuity via a topology change in compact stars,''
Prog. Part. Nucl. Phys. \textbf{113}, 103791 (2020);
``Manifestation of hidden symmetries in baryonic matter: from finite nuclei to neutron stars,''
Mod. Phys. Lett. A \textbf{36},  2130012 (2021).
%doi:10.1142/S0217732321300123


\bi{FR} B.~Friman and M.~Rho,
``From chiral Lagrangians to Landau Fermi liquid theory of nuclear matter,''
Nucl. Phys. A \textbf{606}, 303-319 (1996).

\bi{shankar} R.~Shankar,
``Renormalization group approach to interacting fermions,''
Rev. Mod. Phys. \textbf{66}, 129-192 (1994).

\bi{hu-jain} Y. Hu and J.K. Jain, ``Kohn-Sham theory of fractional quantum Hall effect," Phys. Rev. Lett. {\bf 123}, 176802 (2019).

\bi{lattimer} N.~Rutherford, M.~Mendes, I.~Svensson, A.~Schwenk, A.~L.~Watts, K.~Hebeler, J.~Keller, C.~Prescod-Weinstein, D.~Choudhury and G.~Raaijmakers, \textit{et al.}
``Constraining the dense matter equation of state with new NICER mass-radius measurements and new chiral effective field theory inputs,''
[arXiv:2407.06790 [astro-ph.HE]].

\bi{dlfp-Bira} S.~R.~Beane and U.~van Kolck,
``The dilated chiral quark model,''
Phys. Lett. B \textbf{328}, 137-142 (1994) [arXiv:hep-ph/9401218 [hep-ph]].


\bi{appelquist} T.~Appelquist, J.~Ingoldby and M.~Piai,
``Dilaton Effective Field Theory,''
Universe \textbf{9}, no.1, 10 (2023).

\bi{gA*} Y.~L.~Ma and M.~Rho,
``Quenched $g_A$ in nuclei and emergent scale symmetry in baryonic matter,''
Phys. Rev. Lett. \textbf{125}, no.14, 142501 (2020)


\bi{canfora} E.g. among many in the literature,  F.~Canfora, E.~Delgado and L.~Urrutia,
``Ordered patterns of (3+1)-dimensional hadronic gauged solitons in the low-energy limit of Quantum Chromodynamics at a finite baryon density, Their magnetic fields and novel BPS bounds,'' Symmetry \textbf{16}, no.5, 518 (2024);  M.~Eto, K.~Nishimura and M.~Nitta,
``How baryons appear in low-energy QCD: Domain-wall Skyrmion phase in strong magnetic fields,''
[arXiv:2304.02940 [hep-ph]].

\bi{paeng-park-vento} B.~Y.~Park, W.~G.~Paeng and V.~Vento,
``The inhomogeneous phase of dense skyrmion matter,''
Nucl. Phys. A \textbf{989}, 231-245 (2019) [arXiv:1904.04483 [hep-ph]].

\bi{MR-fractional} M.~Rho,
``Fractionalized quasiparticles in dense baryonic matter,''
[arXiv:2004.09082 [nucl-th]]; 
``Probing fractional Quantum Hall sheets in dense baryonic matter,''
[arXiv:2211.14890 [nucl-th]].

\bi{cho} Y.~M.~Cho, K.~Kimm, J.~H.~Yoon and P.~Zhang,
``New topological structures of Skyrme theory: Baryon number and monopole number,''
Eur. Phys. J. C \textbf{77}, no.2, 88 (2017); P.~Zhang, K.~Kimm, L.~Zou and Y.~M.~Cho,
``Re-interpretation of Skyrme theory: New topological structures,''
[arXiv:1704.05975 [hep-th]].

\bi{QCP} T.~Senthil, A.~Vishwanath, L.~Balents, S.~Sachdev and M.~P.~A.~Fisher,
``Deconfined quantum critical Point,'' Science \textbf{303}, no.5663, 1490-1494 (2004).

\bi{magic} G.~E.~Brown, M.~Rho and W.~Weise,
``Phenomenological sizes of confinement regions in baryons,''
Z. Phys. A \textbf{331}, 139-149 (1988).

\bi{domain-wall} T.~Sulejmanpasic, H.~Shao, A.~Sandvik and M.~Unsal,
``Confinement in the bulk, deconfinement on the wall: infrared equivalence between compactified QCD and quantum magnets,''
Phys. Rev. Lett. \textbf{119}, no.9, 091601 (2017).

\bi{evidence-core} E.~Annala, T.~Gorda, A.~Kurkela, J.~N\"attil\"a and A.~Vuorinen,
``Evidence for quark-matter cores in massive neutron stars,''
Nature Phys. \textbf{16}, no.9, 907-910 (2020).

\bi{CFL} M.~G.~Alford, K.~Rajagopal and F.~Wilczek,
``Color flavor locking and chiral symmetry breaking in high density QCD,''
Nucl. Phys. B \textbf{537}, 443-458 (1999);
 T.~Sch\"afer and F.~Wilczek,``Continuity of quark and hadron matter,'' Phys. Rev. Lett. \textbf{82}, 3956-3959 (1999).

\bi{kaplan}D.~B.~Kaplan,
``Qualitons,'' Nucl. Phys. B \textbf{351}, 137-160 (1991).

\bi{superqualiton} D.~K.~Hong, M.~Rho and I.~Zahed,
``Qualitons at high density,''
Phys. Lett. B \textbf{468}, 261-269 (1999).

\bi{QHL} K.~Nishimura, N.~Yamamoto and R.~Yokokura,
``Quantum Hall liquids in high-density QCD,''
[arXiv:2410.07665 [hep-th]].

\end{thebibliography}
\end{document}